\documentstyle[12pt]{article}
\begin{document}
\begin{center}
{\Large \bf
The extra high energy cosmic rays spectrum
in view of the decay of proton at the Planck scale
}
\bigskip

{\large D.L.~Khokhlov}
\smallskip

{\it Sumy State University, R.-Korsakov St. 2\\
Sumy 244007 Ukraine\\
e-mail: khokhlov@cafe.sumy.ua}
\end{center}

\begin{abstract}
The structure of the extra high energy cosmic rays spectrum
in view of the decay of proton is considered.
The time required for proton travel from the source to the
earth defines the limiting energy of proton.
Protons with the energies more than the limiting energy decay and
do not give contribution in the EHECRs spectrum.
It is assumed that proton decays at the Planck scale.
Depending on the range of distances to the EHECRs sources,
the range of the limiting energies of proton is determined.
This allows one to explain the structure of the EHECRs spectrum.
\end{abstract}

The energy spectrum of extra high energy cosmic rays
(EHECRs) above $10^{10}\ {\rm eV}$ can be
divided into three regions: two "knees" and one "ankle"~\cite{Yo}.
The first "knee" appears around $3\times 10^{15}\ {\rm eV}$
where the spectral power law index changes from $-2.7$ to $-3.0$.
The second "knee" is somewhere between $10^{17}\ {\rm eV}$ and
$10^{18}\ {\rm eV}$ where the spectral slope steepens from
$-3.0$ to around $-3.3$. The "ankle" is seen in the region of
$3 \times 10^{18}\ {\rm eV}$ above which the spectral slope
flattens out to about $-2.7$.

Consider the structure of the EHECRs spectrum
in view of the decay of proton.
The lifetime of proton relative to the decay of proton
at the scale of the mass $M$
is given by
\begin{equation}
t_p=\frac{M^4}{E^5}.
\label{eq:tp}
\end{equation}
From this the time required for proton travel from the source to the
earth defines the limiting energy of proton
\begin{equation}
E_{lim}=\left(\frac{M^4}{t}\right)^{1/5}.
\label{eq:E}
\end{equation}
Within the time $t$, protons
with the energies more than the limiting energy $E>E_{lim}$ decay and
do not give contribution in the EHECRs spectrum.

Let us assume that proton decays
at the Planck scale $m_{Pl}=1.2 \times 10^{19}\ {\rm GeV}$.
Determine
the range of the limiting energies of proton
depending on the range of distances to the EHECRs sources.
Take the maximum and minimum distances to the source as
the size of the universe and the thickness
of our galactic disc respectively.
For the lifetime of the universe
$t_U=1.06 \times 10^{18}\ {\rm s}$~\cite{Kh1},
the limiting energy is equal to $E_U=6.7 \times 10^{15}\ {\rm eV}$.
This corresponds to the first "knee" in the EHECRs spectrum.
For the thickness of our galactic disc $\simeq 300\ {\rm pc}$,
the limiting energy is equal to $E_G=1.1 \times 10^{18}\ {\rm eV}$.
This corresponds to the second "knee" in the EHECRs spectrum.
Thus
the range of the limiting energies of proton
due to the decay of proton at the Planck scale
lies between
the first "knee" $E\sim 3\times 10^{15}\ {\rm eV}$ and
the second "knee" $E\sim 10^{17}-10^{18}\ {\rm eV}$.

From the above consideration it follows that
the decrease of the spectral power law index from $-2.7$ to $-3.0$
at the first "knee" $E\sim 3\times 10^{15}\ {\rm eV}$ and
from $-3.0$ to around $-3.3$
at the second "knee" $E\sim 10^{17}-10^{18}\ {\rm eV}$
can be explained as a result of
the decay of proton at the Planck scale.
From this it seems natural that, below
the "ankle" $E<3 \times 10^{18}\ {\rm eV}$,
the EHECRs events are mainly caused by the protons.
Above the "ankle" $E>3 \times 10^{18}\ {\rm eV}$,
the EHECRs events are caused by the particles other than protons.

\end{document}